\documentclass{IEEEmce}

\usepackage[colorlinks,urlcolor=blue,linkcolor=blue,citecolor=blue]{hyperref}
\usepackage{rotating}
\usepackage{graphicx}
\usepackage{caption}  
\usepackage{pdflscape}
\usepackage{placeins}

\usepackage{array}
\usepackage{booktabs}
 
\usepackage{pgfplots}
\usepackage{pgfplotstable}
\pgfplotsset{compat=1.18}

\jvol{XX}

\begin{document}

\sptitle{Native and Cross-Platform Mobile Development}


\title{Cross-Platform vs Native Mobile Development: An Empirical Study of Software Quality Trade-offs}


\author{Alexandru Ilovan}

\begin{abstract}
Cross-platform mobile frameworks promise code reuse, shorter delivery cycles, and lower implementation effort, but their trade-offs relative to native development remain difficult to assess objectively. Many comparisons rely on simplified applications, inconsistent feature sets, or a narrow set of metrics. This paper compares five implementations of the same plant-management application: native iOS, native Android, Flutter, React Native, and Kotlin Multiplatform. The shared approaches target both Android and iOS, yielding eight executable variants. Guided by ISO/IEC 25010, the study examines time behavior, implementation footprint, source-code organization, and observable rendering responsiveness.

All implementations share the same domain, backend services, functional requirements, and benchmark contract. The supplied dataset contains 2,000 completed runs per variant and covers authenticated and cached retrieval, image transfer and decoding, list rendering and scrolling, local synchronization, and media upload. Backend preparation and framework-specific UI drivers are analyzed separately from the primary client workflow.

Native records the lowest non-UI client subtotal on both operating systems, with Kotlin Multiplatform the closest cross-platform implementation, while operation and UI-wrapper rankings vary by task. The shared approaches contain less authored mobile source than the native applications combined, and the source inventory shows different patterns of file size, organization, and dependency use. Rather than identifying a universally superior technology, the study shows that each approach's advantages and costs depend on the quality attribute, workload, platform, and measurement boundary.

\hfill


\hfill

\end{abstract}

\maketitle

\enlargethispage{10pt}

\chapterinitial{Introduction}

Mobile software teams frequently need to deliver the same product on Android and iOS while preserving acceptable runtime behavior, interface responsiveness, maintainability, and engineering cost. Native development provides direct access to platform APIs, vendor tooling, and platform-specific conventions, but separate native codebases can duplicate implementation and maintenance work. Cross-platform frameworks reduce some of that duplication through shared source code and common development workflows, while introducing their own runtimes, rendering models, dependencies, and platform-integration boundaries.

The resulting choice cannot be reduced to a universal framework ranking because the relevant quality dimensions measure different properties. Fast execution does not imply low energy use, a compact source tree does not establish short development time, and static-analysis grades do not directly demonstrate long-term maintainability. Observable rendering timings also cannot substitute for subjective user-experience evidence. A rigorous comparison must therefore define each construct, apply equivalent workloads, and limit conclusions to the quantities that are actually measured.

Comparability is the central methodological challenge. Differences attributed to a framework may instead arise from application scope, backend behavior, hardware, operating system, framework version, optimization choices, or measurement technique. This study responds with a non-trivial application, a shared backend, repeated execution, and a common telemetry contract. It compares Flutter, React Native, and Kotlin Multiplatform with native Android and native iOS baselines. Runtime observations are compared within each operating system, while source properties are treated at codebase level rather than duplicated for every executable target.

The main contribution is a common evaluation framework that replaces informal comparisons with measurable, bounded evidence.

\subsection{Scope and Contributions}

The present study addresses gaps associated with simplified benchmark applications, inconsistent feature sets, and narrow metric coverage through five implemented codebases and eight measured runtime variants of a common mobile application. It compares native and cross-platform approaches on Android and iOS using a shared benchmark contract, runtime telemetry, and source-code analysis. The scope combines within-platform runtime comparisons with structural assessments, covering time behavior, rendering-responsiveness proxies, implementation footprint, file-level source concentration, and declared dependency surfaces.

The experimental artifact used for this comparison is \textit{ChloroByte}, a plant management app created for this study. It comprises native iOS with Swift and SwiftUI, native Android with Kotlin and Jetpack Compose, Flutter with Dart, React Native with TypeScript and Expo, and Kotlin Multiplatform with shared Kotlin services and native interfaces. Flutter, React Native, and Kotlin Multiplatform each target Android and iOS, yielding eight runtime variants. The common application contract includes login and session restoration, remote API communication, plant list and detail flows, search, filtering and sorting, image loading and upload, local persistence, settings management, and an integrated benchmark interface. The complete repositories retain some platform-specific differences, so runtime claims are bounded to the scripted benchmark path rather than assuming whole-product equivalence.

Each implementation contains a benchmark runner that executes the same nominal 12-step scenario and records structured telemetry. Backend reset and dataset seeding are instrumented preparation steps. They contribute to whole-run elapsed time but are interpreted separately from operations that exercise authentication, plant retrieval, cache validation, image transfer and decoding, list rendering, automated scrolling, local synchronization, and upload. The supplied exports provide all wrapper totals and aggregate operation summaries for 2,000 runs per variant. They retain complete nested telemetry only for the last run, which limits distributional and inferential operation-level analysis.

The ISO/IEC 25010 product quality model informs the placement of the measured runtime and structural properties within a broader software-quality context.\footnote{ISO/IEC 25010 product quality model: \url{https://iso25000.com/index.php/en/iso-25000-standards/iso-25010}.} The empirical questions are deliberately narrower than the full quality model. They concern measured time behavior, observable rendering-responsiveness proxies, implementation footprint, file-level source concentration, and declared dependency surfaces. This boundary keeps the conclusions aligned with the quantities recorded in the runtime exports and source inventory.

The contribution is threefold. First, the study documents five implementations and eight platform-specific runtime variants of the same application, including a Kotlin Multiplatform design that shares non-UI application layers while retaining native UIs. Second, it provides a repeated telemetry workflow and a 16,000-run dataset for comparing native, Flutter, React Native, and Kotlin Multiplatform within the same operating system. Third, it reports client-operation timing, rendering wrappers, authored source footprint, file-level source concentration, and declared dependency surfaces as distinct measures. This separation supports bounded framework selection without assuming that one technology must dominate every quality attribute.

\subsection{Organization of the Paper}

The remainder of the paper is organized as follows. Background Work synthesizes prior research on framework selection, runtime behavior, energy, maintainability, and code reuse, leading to the Research Questions. Methodology defines the study design, units of analysis, operational measures, inclusion rules, and analysis boundaries. Artifact and Implementation Design describes the five codebases, benchmark instrumentation, and parity limitations. Experimental Setup and Protocol records the backend, workload, supplied run organization, repository-measurement rules, and reproducibility gaps. Results present dataset validity and findings by research question. Discussion relates the findings to prior work and practical framework selection, followed by Limitations and Conclusions and Future Work.

\section{Background work}

Prior research approaches mobile-framework choice through architecture taxonomies, selection criteria, practitioner and ecosystem evidence, controlled runtime experiments, energy measurements, source-code analysis, and application case studies. These forms of evidence address related but distinct questions. Their conclusions can be compared only within the applications, platforms, metrics, and procedures from which they were obtained.

Ecosystem, runtime, and source-quality studies illustrate the separation among forms of evidence. Jo\v{s}t and Taneski combine surveys, repository activity, search interest, online discussion, and job postings for Flutter, React Native, and .NET MAUI \cite{Jost2025StateOfArtCrossPlatformMobileApplicationDevelopmentFrameworksComparativeStudyOfMarketAndDeveloperTrends}. Flutter leads some admired or loved indicators, while React Native is more visible in one employment dataset, yet several direct popularity differences are not statistically significant. Bedogni et al. instead measure latency in six operating-system, framework, and mapping-SDK combinations \cite{Bedogni2025OnLatencyPerformanceOfMobileMappingServicesTowardsVulnerableRoadUsersSafety}. Native responds faster for in-application alerts, but Flutter with Mapbox is faster than Kotlin with Mapbox for some Android map operations. Ugli et al. move from runtime to source quality by applying SonarQube and Code Climate to controlled and repository applications \cite{Ugli2025CodeSmellsDevelopmentEfficiencyNativeCrossPlatformMobileAppsStudyOfJavaKotlinFlutter}. Java produces more raw smells in most controlled comparisons and Dart receives strong maintainability grades. Together, these studies show that adoption, latency, and analyzer findings cannot be combined into one framework score.

Selection-oriented comparisons make criteria explicit but provide less direct experimental evidence. Zou and Darus compare five frameworks across seven qualitative dimensions, while Adnane et al. convert nine equally weighted binary criteria into a developer recommendation tool \cite{Zou2024ComparativeAnalysisCrossPlatformMobileDevelopmentFrameworks,Souha2024ComparativeAnalysisMobileApplicationFrameworksDeveloperGuideChoosingRightTool}. Their explicit criteria help reveal what a decision model values, but binary scoring and equal weighting obscure context and interaction among criteria. Their categories also differ from other surveys, and Adnane et al. include SwiftUI, a native Apple UI technology, among alternatives framed for the same decision. Compared with the preceding empirical studies, these papers are better suited to eliciting requirements than to establishing runtime or maintainability rankings.

Controlled runtime and energy experiments provide the broadest range of empirical designs. Oliveira et al. use computational benchmarks and interaction-oriented applications and find native Android generally the most consistently resource efficient, with Flutter often the strongest cross-platform option for CPU-intensive work \cite{Oliveira2023AnalyzingResourceUsageOverheadMobileAppDevelopmentFrameworks}. React Native improves when animation is delegated to its native driver, whereas Ionic performs poorly in several animation and scrolling cases. Frattaroli et al. compare complete movie applications built with native Swift, native Kotlin, Kotlin Multiplatform Mobile, Flutter, and React Native \cite{Frattaroli2023EcologicalImpactNativevsCrossPlatformMobilePreliminaryStudy}. Native produces the smallest packages, while network and energy results vary by operating system and some iOS quantities, including Kotlin Multiplatform energy, are unavailable. Huber et al. obtain stronger energy isolation by measuring 14 Android UI-component implementations with external hardware \cite{Huber2023OnEnergyEfficiencyOfHybridUIComponentsForMobileCrossPlatformDevelopment}. Native components are consistently most efficient, but the ordering among wrapped web components and Capacitor changes across dialogs, sheets, drawers, and scrolling. Oliveira et al. provide workload breadth, Frattaroli et al. provide cross-OS application breadth, and Huber et al. provide component-level instrumentation. Their different winners reinforce that the unit of work and measurement boundary determine the conclusion.

Other studies from the same period examine source quality, organizational priorities, and implementation feasibility. Karami et al. combine a systematic review of 75 studies with 3,566 native Android and React Native repositories and report fewer detected smells for React Native in small and medium project groups \cite{Karami2023OnImpactOfDevelopmentFrameeworksOnMobileApps}. This breadth complements Ugli et al.'s tighter controlled comparison, but unrelated repositories and language-specific rule sets prevent a causal claim about framework maintainability. Mahmoud et al. survey 85 practitioners and report strong Flutter adoption together with memory and speed concerns; the convenience sample is concentrated among young developers, Egyptian respondents, and small organizations \cite{Mahmoud2023IndustrialPractictionerPerspectiveMobileApplicationsProgrammingLanguagesSystems}. El Tom et al. apply 33 criteria, expanded into as many as 162 checklist items, to React Native and Xamarin.Forms in one organization and score React Native higher, although package size is the principal directly measured performance quantity \cite{ElTom2023CriteriaBasedEvaluationCrossPlatformDevelopmentFrameworks}. These studies expose real decision criteria but do not measure functionally equivalent applications under common controls.

Application case studies clarify where reuse creates integration work. De Almeida et al. report that adapting the Flutter-based IscteSpots application to the web changes 27 of 155 Dart files and 2,115 of 14,309 lines, partly because mobile packages lack web support \cite{DeAlmeida2023CrossPlatformMobileAppDevelopmentTheISCTESpotsExperience}. Seo embeds Unity3D in a React Native storybook application and records 825 completed tasks after preliminary fixes, demonstrating feasibility while introducing communication, permission, state, and memory-management complexity \cite{Seo2023CaseStudyCombiningTwoCrossPlatformDevelopmentFrameworksStorybookMobileApp}. Zarichuk provides a broader comparison of native, hybrid, and cross-platform approaches, but classifies React Native differently from studies that distinguish native-UI runtimes from WebView-based hybrids \cite{Zarichuk2023ComparativeAnalysisFrameworksMobileApplicationDevelopmentNativeHybridCrossPlatformSolutions}. These papers show that shared source moves effort toward packages, adapters, and platform integration rather than eliminating platform-specific work.

Architecture and reuse research continues the contrast between classification and measured sharing. Stanojevi\'c et al. survey modern cross-platform architectures and selection considerations, providing a useful map of approaches but no controlled application comparison \cite{Stajojevic2022OverviewModernCrossPlatformMobileFrameworks}. Wheeler and Olszewska report 69.16\% shared code in a custom C++ framework for smart services, but exclude boilerplate and UI layout from the denominator \cite{Wheeler2022CrossPlatformMobileApplicationDevelopmentForSmartServices}. The first study broadens the design space, while the second quantifies reuse within one particular sharing boundary. Neither percentage nor architecture category directly establishes runtime quality or development effort.

Broader framework evaluations likewise depend on scope. Nawrocki et al. compare native Android, native iOS, Xamarin.Forms, React Native, and Flutter, finding native advantages in package size and startup and identifying Flutter as the strongest overall cross-platform compromise in their workload \cite{Nawrocki2021ComparisonOfNativeAndCrossPlatformFrameworksMobileApplications}. One device per operating system, synthetic applications, partly manual measurements, and no inferential analysis limit that ordering. Blanco and Lucr\'edio reduce the counted implementation from 3,719 native lines to 885 model-language and native lines through a holistic model-driven approach, but exclude view code and do not evaluate runtime behavior \cite{Blanco2021HolisticApproachCrossPlatformSoftwareDevelopment}. Shetty and Padmashree compare development approaches and framework characteristics at a broader level without an equivalent repeated artifact \cite{Shetty2021EvaluationCrossPlatformApplicationDevelopmentFrameworks}. Their results are useful for package size, startup, and potential reuse, but those constructs cannot stand in for overall software quality.

Controlled device and list-render experiments also show unstable rankings. Bi{\o}rn-Hansen et al. conduct an Android experiment across six physical devices and 16,290 observations \cite{Biorn-Hansen2020EmpiricalInvestigationCrossPlatformMobileDevelopmentFrameworks}. Native Android performs best overall, yet NativeScript leads parts of the completion-time and CPU comparisons, and Flutter combines low incremental memory use with high idle memory. I\c{s}{\i}tan and K\"okl\"u test a 1,000-item list on one Android emulator without a native baseline or reported repetitions and observe the shortest render time for NativeScript \cite{Isitan2020ComparisonEvaluationCrossPlatformMobileApplicationDevelopmentTools}. The larger study offers stronger device and observation coverage, while the narrower study demonstrates how virtualization and a single task can change the apparent leader.

Across the literature, native implementations frequently lead package-size, startup, device-feature, and component-level energy measures, but cross-platform approaches remain competitive for operations aligned with their compilation, rendering, or native-delegation mechanisms. Popularity, code sharing, latency, memory, energy, and source-quality measures answer different questions. Four gaps remain especially relevant. Few studies combine feature-aligned native, Flutter, React Native, and Kotlin Multiplatform applications on both operating systems. Repeated application workflows with operation-level telemetry are uncommon. Rendering-wrapper times are often discussed without common frame traces or subjective evaluation. Source volume and code-sharing percentages are also prone to interpretation without an explicit counting scope.

The present study addresses these gaps with five implemented codebases and eight measured variants. Its shared workflow exercises authentication, remote access, cache validation, image handling, local persistence, list rendering, scrolling, and upload. Runtime comparisons remain within operating system, while the source inventory applies one stated counting scope to all five codebases. The research questions therefore distinguish time behavior, observable rendering evidence, total implementation footprint, and file-level source and dependency structure.

\section{Research Questions}

The literature shows that broad framework rankings conceal workload effects, operating-system differences, and measurement limitations. Accordingly, this study does not seek a universally superior technology. It asks four bounded questions across the five implemented codebases.

\noindent\textbf{RQ1. Within each target operating system, how do native, Flutter, React Native, and Kotlin Multiplatform builds differ in client-workflow elapsed time and operation-level time behavior under the common repeated benchmark workflow?}

RQ1 compares eight runtime variants: native iOS, Flutter iOS, React Native iOS, Kotlin Multiplatform iOS, native Android, Flutter Android, React Native Android, and Kotlin Multiplatform Android. Its primary aggregate is the sum of seven measured non-UI client operations: authentication, list retrieval, conditional retrieval, image transfer, image decoding, local synchronization, and upload. Full-scenario elapsed time is retained as contextual telemetry. Backend reset and seeding are excluded from the client subtotal because they primarily measure shared server and object-storage preparation.

\noindent\textbf{RQ2. Within each target operating system, how do native, Flutter, React Native, and Kotlin Multiplatform builds differ in the recorded list-render wrapper, and how comparable are their automated-scroll wrapper durations?}

RQ2 uses the available render and scroll wrapper summaries as descriptive evidence. The render wrapper measures completion of each framework's scripted synthetic-list task. The scroll wrapper is analyzed as a property of the realized test driver because the implementations use different animation and completion rules. Neither quantity is interpreted as a direct measure of frame quality, perceived smoothness, or subjective user experience.

\noindent\textbf{RQ3. How does authored mobile-source footprint differ among the native iOS, native Android, Flutter, React Native, and Kotlin Multiplatform codebases in the inspected snapshot, and how does each shared codebase compare with the combined native footprint required for both operating systems?}

Implementation footprint is evaluated through authored file counts, physical lines, nonblank lines, and source lines of code. Native iOS and native Android are presented separately and as a combined two-platform strategy. The same inclusion and exclusion rules are applied to each inspected source tree.

\noindent\textbf{RQ4. How do the five codebases differ in the organization of their authored source files and in their use of direct production dependencies?}

RQ4 compares all five codebases using median SLOC per authored file, maximum authored-file SLOC, and direct production dependency declarations. These measures describe how source is distributed across files and how many external dependencies are declared by each inspected implementation under the stated counting rules.

\section{Methodology}

The study is a comparative case study of native and cross-platform mobile development. It contains five source codebases and eight executable variants. Native iOS and native Android each contribute one platform-specific build, while Flutter, React Native, and Kotlin Multiplatform each contribute Android and iOS builds. The runtime phase is a high-repetition observational benchmark of the supplied artifacts.

\subsection{Related-Work Identification}

The Background Work synthesis was assembled through a structured qualitative search whose identification and screening counts are documented using PRISMA 2020 as reporting guidance \cite{PRISMA}. Searches in Web of Science and IEEE Xplore were supplemented with Connected Papers, Perplexity.ai, and Consensus as discovery tools. The seed query was \texttt{mobile cross platform native}. Screening yielded 61 candidate publications, of which 21 were retained for full-text qualitative synthesis because they directly examined native or cross-platform mobile development. The synthesis motivates and delimits the research questions; it does not pool effect estimates across incompatible experiments.

\begin{figure}[htbp]
\centering
\includegraphics[width=0.95\columnwidth,height=0.46\textheight,keepaspectratio]{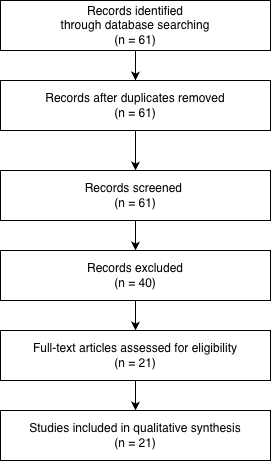}
\caption{PRISMA 2020 flow diagram for identification and selection of studies included in the qualitative literature review.}
\label{fig:review-flow}
\end{figure}

Figure \ref{fig:review-flow} presents the identification, screening, eligibility, and inclusion flow for the literature review.

The diagram documents the reported review flow.

\subsection{Study Design and Units of Analysis}

The study uses two units of analysis because runtime behavior and source-code properties belong to different artifacts. A runtime observation is one completed execution of a specific build on a specific operating system. RQ1 and RQ2 compare four build conditions within Android and four within iOS. Native Android is not compared directly with native iOS because that contrast would combine framework, operating-system, device, and toolchain effects. A codebase observation is one inspected mobile source tree. RQ3 and RQ4 therefore concern five codebases rather than eight executable targets, with an additional combined-native view for RQ3. Table \ref{tab:study-units} summarizes the implemented conditions.

\begin{table*}[t]
\centering
\caption{Implemented study units.}
\begin{tabular}{llll}
\toprule
Approach & Primary source and UI & Android build & iOS build \\
\midrule
Native Android & Kotlin and Jetpack Compose & Available & N/A \\
Native iOS & Swift and SwiftUI & N/A & Available \\
Flutter & Dart and Flutter & Available & Available \\
React Native & TypeScript and Expo & Available & Available \\
Kotlin Multiplatform & Shared Kotlin and native UI shells & Available & Available \\
\bottomrule
\end{tabular}
\label{tab:study-units}
\end{table*}

The mobile applications use a common plant-management domain and expose aligned benchmark entry points. All benchmark clients authenticate against the same Node.js backend, use the same logical benchmark account and API contract, request the same configured dataset, and execute the same nominal sequence. Idiomatic framework choices are permitted. A measured difference is interpreted as a framework-level comparison only to the extent that inputs, outputs, and timing boundaries are equivalent.

\subsection{Functional Equivalence and Experimental Controls}

Functional equivalence is assessed at two levels. The first level is the user-visible application contract, including authentication, plant retrieval, caching, image handling, local persistence, and list interaction. The second level is the benchmark contract, which specifies the ordered logical operations, input sizes, telemetry names, outputs, and success conditions. Equivalence therefore concerns observable behavior and workload rather than an identical sequence of internal calls. Requiring identical internals would be impractical because each platform and framework exposes different APIs, execution models, and integration mechanisms. Platform-specific product features outside the common benchmark path are excluded from runtime comparisons and are considered separately in the repository inventory.

Many internal differences follow directly from platform constraints and framework-specific APIs. Native applications use operating-system networking, security, persistence, and image-decoding facilities. Flutter reaches corresponding services through its engine and plugin abstractions, while React Native and Expo delegate work through JavaScript-to-native interfaces and native-backed library objects. Kotlin Multiplatform shares application logic but supplies platform-specific implementations for networking, secure storage, databases, image decoding, and UI. Consequently, the same logical operation can involve different bridges, object representations, scheduling behavior, and library calls.

Same-operating-system comparisons reduce operating-system confounding, but they cannot separate the effects of the framework, supporting libraries, platform APIs, architecture, and implementation boundaries. The recorded durations therefore characterize the complete realized implementations under a common logical workflow. They do not estimate an isolated causal effect of framework overhead. Backend reset and seed operations are treated separately because they measure shared server and object-storage preparation rather than client-framework behavior.

\subsection{Operationalization of the Research Questions}

Table \ref{tab:rq-operationalization} maps each research question to its unit, measures, and interpretation boundary.

For RQ1, the primary non-UI client subtotal is the sum of the exported means for \texttt{step\_login}, \texttt{step\_fetch\_plants}, \texttt{step\_fetch\_plants\_304}, \texttt{step\_download\_images}, \texttt{step\_decode\_image}, \texttt{step\_sqlite\_sync}, and \texttt{step\_upload\_image}. The subtotal excludes the runtime-origin marker, backend reset and seed, rendering and scrolling, and uninstrumented orchestration gaps.

\begin{table*}[t]
\centering
\caption{Operationalization of the research questions.}
\begin{tabular}{p{0.07\textwidth}p{0.21\textwidth}p{0.30\textwidth}p{0.31\textwidth}}
\toprule
RQ & Unit of analysis & Measures & Interpretation boundary \\
\midrule
RQ1 & One completed execution; eight build conditions & Non-UI client subtotal, operation means, and contextual full-scenario duration & Same-OS comparison; reset and seed are backend preparation \\
RQ2 & One wrapper observation; eight build conditions & Synthetic-list render and automated-scroll wrapper durations & Drivers differ; no frame-quality or subjective UX claim \\
RQ3 & One inspected source tree; five codebases & Authored files, PLOC, nonblank lines, SLOC, and combined-native total & Common counting scope across the inspected snapshot \\
RQ4 & Five inspected source trees & Median and maximum file SLOC; direct production dependency declarations & File structure and package granularity differ \\
\bottomrule
\end{tabular}
\label{tab:rq-operationalization}
\end{table*}

For RQ2, the runners emit list-render and automated-scroll wrappers, with some nested first-visible and frame-related events. The exports retain those detailed nested events only for the final run, and their semantics differ across frameworks. Flutter and iOS use fixed-duration animations, Android Compose uses framework-controlled movement with fallback behavior, and React Native completes its driver substantially sooner. The available aggregate wrappers are reported as implementation and harness profiles.

For RQ3, the inspected-source inventory includes authored production-language files in each mobile source tree and excludes tests, dependencies, generated output, assets, caches, and exported telemetry. Benchmark and telemetry source remains included because it is intertwined with application shells in some implementations. Kotlin Multiplatform is separated into \texttt{commonMain}, \texttt{androidMain}, \texttt{iosMain}, and native UI shells. The shared-source ratio does not duplicate \texttt{commonMain}. Native iOS and native Android are reported separately and as their sum. The backend is shared infrastructure and is excluded from client totals.

For RQ4, SLOC is aggregated per included authored file using the same lexical counting rules as RQ3. The median describes the central file size; for an even number of files, it is the mean of the two central values. The maximum identifies the largest included source file in each codebase. Direct production dependencies are counted from the maintained project manifests. Internal project edges, test dependencies, development-only tools, build-time processors, framework-managed SDK entries, and platform bill-of-materials constraints are excluded.

\subsection{Data Inclusion and Analysis Plan}

The wrapper-level inclusion rule requires \texttt{succeeded=true} and \texttt{completedSteps=totalSteps=12}. All 16,000 supplied wrapper rows meet that rule. The operation-summary tables also report 2,000 samples for every step.

For full-scenario elapsed time, the analysis reports sample size, mean, sample standard deviation, median, 95th percentile, minimum, and maximum from the 2,000 wrapper rows. Percentiles use the linearly interpolated Type-7 sample-quantile estimator. Operation-level reporting uses the exported mean and sample standard deviation. Relative differences are calculated within operating system as $100(\bar{x}_{candidate}/\bar{x}_{native}-1)$.

\section{Artifact and Implementation Design}

The experimental artifact comprises five mobile clients, the shared backend, and embedded benchmark and telemetry subsystems. The design goal is functional alignment at the experimental boundary while allowing each application to use framework-appropriate architecture internally. This section describes the inspected working trees and the boundaries that affect interpretation.

\subsection{Shared Functional Contract and Architecture}

The common experimental path includes authentication, bearer-token refresh, plant retrieval, HTTP cache validation, signed image access, image upload, synthetic list rendering, local write and read operations, telemetry storage, and CSV export. Each client uses a central dependency container or equivalent composition root and separates API, authentication, persistence, feature, benchmark, and telemetry responsibilities. The clients use platform-specific mechanisms to route or rewrite loopback image URLs to a configured backend host.

\subsection{Native Baselines}

\subsubsection{Native iOS}

The current iOS application uses SwiftUI with Observation-based view models and a central \texttt{AppContainer}. A shared API client supports the normal product and benchmark paths. The product repository performs authenticated REST retrieval, CRUD operations, signed-image access, and image upload, and it persists a GRDB plant cache. GRDB also stores telemetry and benchmark records, Keychain stores access and refresh credentials, and an ephemeral \texttt{URLSession} configuration disables URL caching and cookie storage for the benchmark network client. The project uses NukeUI for image presentation.

\subsubsection{Native Android}

The Android application uses Kotlin, Jetpack Compose, \texttt{ViewModel}, \texttt{StateFlow}, and Navigation Compose. Room provides SQLite persistence; credentials are stored with encrypted shared preferences; OkHttp performs authenticated API communication; and Coil handles remote images. The project targets SDK 35, uses Java 17, and leaves release minification disabled in the inspected configuration. The benchmark is part of the main source set and can be launched from the normal application route or a benchmark-only intent. The current network-security configuration permits the configured local benchmark hosts.

\subsection{Flutter}

The Flutter implementation uses Dart, Material widgets, a central application container, and \texttt{ChangeNotifier}-based feature controllers. It uses the \texttt{http} package for networking, \texttt{sqflite} for local database operations, and \texttt{flutter\_secure\_storage} for credentials. Android and iOS runner projects are present. The repository declares Flutter and Dart constraints and includes platform wrappers generated by Flutter tooling; generated wrappers will not be counted as authored product source.

\subsection{React Native}

The React Native implementation uses TypeScript, Expo, React hooks, and a central service container. Manual route state is used instead of an external navigation package. Expo Secure Store persists credentials, while a shared SQLite database accessed through Expo SQLite stores plant-cache, telemetry, and benchmark records. The local synchronization benchmark transactionally writes and reads 200 deterministic records, with the workload and timing boundaries aligned with native iOS: write timing includes commit, and read timing includes record materialization. Expo File System, Image Picker, and Sharing support media and result-export workflows. The checked-in configuration uses Expo 55, React Native 0.83, React 19, TypeScript 5.9, Hermes, and the React Native new architecture.

\subsection{Kotlin Multiplatform}

The Kotlin Multiplatform implementation shares application services rather than UI code. \texttt{commonMain} contains models and data transfer objects, sorting, authentication and refresh handling, the plant repository, Ktor API access, ETag and JSON caching, SQLDelight persistence, image transfer, telemetry, CSV export, and the 12-step benchmark runner. \texttt{androidMain} supplies the Ktor OkHttp engine, Android SQLDelight driver, AES-GCM credential protection backed by Android Keystore, a monotonic clock, and bitmap decoding. Its application shell uses Jetpack Compose. \texttt{iosMain} supplies the Ktor Darwin engine, native SQLDelight driver, Keychain storage, a monotonic clock, and UIKit image decoding. Shared Kotlin is exposed as a static framework to a SwiftUI shell for device and simulator architectures.

The inspected project uses Kotlin 2.0.0, Android Gradle Plugin 8.10.0, Ktor 2.3.12, SQLDelight 2.0.2, coroutines 1.8.1, serialization 1.7.1, datetime 0.6.0, Compose UI 1.5.4, and Material 3 1.1.2. Android targets SDK 35 with minimum SDK 33 and Java 17. The iOS target has deployment version 16.0 and a Swift 5.0 project setting. Both shells implement authentication, cache-backed plant lists, local and server search, sorting and filtering, pagination, detail and CRUD flows, image selection and upload, settings, logout, and benchmark export.

\subsection{Benchmark and Telemetry Instrumentation}

All five clients implement the same nominal sequence of 12 steps: a runtime-origin marker, login, backend reset, backend seed, plant fetch, ETag-based fetch expecting HTTP 304, image download, image decode, list rendering, automated scrolling, local persistence synchronization, and image upload. Default controls expose a short verification run and configurable repeated batches. The supplied configuration contains 2,000 executions per build. Full-scenario elapsed time is recorded by an outer wrapper and is not reconstructed by summing step means because orchestration gaps and instrumentation also contribute.

Telemetry events share a logical schema containing run identifier, timestamp, scenario, operation, duration, success, optional HTTP status, byte counts, and contextual attributes. Each implementation produces a detailed-event table, a per-run wrapper table, and a per-operation aggregate table. All clients use SQLite-backed persistence through GRDB, Room, sqflite, Expo SQLite, or SQLDelight.

\subsection{Functional Parity and Known Implementation Deviations}

The repository audit identified the following deviations that bound the results:

\begin{itemize}
    \item Image wrappers independently acquire their inputs rather than passing the preceding download bytes directly to decoding. Platform decoders differ. React Native times Expo Image's complete \texttt{loadAsync} call and receives a native-backed image reference, rather than performing the same explicit platform-codec operation as the other clients. Upload wrappers also differ in temporary-file and multipart handling.
    \item Synthetic list containers, virtualization, scrolling animation, timeout behavior, and visibility callbacks differ across UI frameworks.
    \item The post-first-row rendering proxies use either a fixed 16 ms delay or one additional rendered frame rather than a measured quiescence condition.
    \item Raw exports cover only the final run even though aggregate tables cover all 2,000 executions.
    \item Runtime-origin markers begin at different lifecycle points and are reused or accumulated throughout an in-process batch, so they do not represent repeated cold starts.
    \item Whole-product scope remains unequal in some media acquisition and platform UI flows. KMP uses native UI shells and therefore represents a different sharing boundary from Flutter and React Native shared-UI approaches.
\end{itemize}

These deviations are properties of the realized artifacts. Results are reported by operation and operating system, and claims are limited to the work performed inside each recorded boundary.

\section{Experimental Setup and Protocol}

The repositories encode a common workload and the supplied CSVs preserve its wrapper and aggregate results. They do not preserve every element needed to reconstruct the execution environment. This section distinguishes verified artifact properties from missing run metadata.

\subsection{Hardware and Software Environment}

The experiments were designed for execution on physical mobile devices. The following mobile devices were used:
\begin{itemize}
    \item iOS device: iPhone 15, iOS version 26.3.1 for the original measurements and 26.6.1 for the React Native rerun;
    \item Android device: Pixel 6 Pro, Android version 16;
\end{itemize}

The experiments were run on Android version 16 for native Android and iOS version 26.3.1 for native iOS; Flutter packages including \texttt{http}, \texttt{sqflite}, and secure storage; Expo 55, React Native 0.83, React 19, TypeScript 5.9, Hermes, and the new architecture for React Native; SwiftUI, GRDB, and NukeUI for native iOS; and the KMP versions reported above.

\subsection{Backend, Network, and Dataset Conditions}

The shared backend is a Fastify and TypeScript service backed by PostgreSQL and S3-compatible object storage. Benchmark routes are available only outside backend production mode and require an authenticated user. \texttt{POST /benchmark/reset} first requests deletion of the authenticated user's distinct image objects. If the storage request throws, the route returns HTTP 503 before deleting database rows. It then deletes the user's plants. \texttt{POST /benchmark/seed} uploads deterministic shared image objects and creates the requested plant and image records sequentially without a transaction or rollback, so a later failure can leave partial state.

The common client configuration requests 200 plants, original and generated images, the \texttt{medium} image variant, a page size of 50, offset zero, and strict failure handling. The medium seed image is a deterministic 256 by 256 noise PNG. In the inspected backend snapshot it is 224,668 bytes. The route uploads one shared original object and one shared generated object, approximately 0.43 MiB in total, then creates 200 plants and associated image records that reference those keys. This work belongs to environment preparation rather than client-framework performance. Seed is omitted from all operation charts and from the primary client subtotal.

\subsection{Benchmark Configuration and Scenario}

The shared workload uses the following values:

\begin{itemize}
    \item benchmark account shared across builds;
    \item 200 seeded plants;
    \item original and generated images enabled;
    \item deterministic medium image variant at 256 by 256 pixels;
    \item page size 50 and offset zero;
    \item strict mode enabled;
    \item a synthetic 50-row render task, a 200-row scroll task, and a 200-record local synchronization task.
\end{itemize}

The 12 wrapper steps execute in this order:

\begin{enumerate}
    \item record the runtime-origin marker;
    \item authenticate the benchmark account;
    \item reset server-side benchmark plants and image objects;
    \item seed the 200-plant dataset;
    \item retrieve one page of plants;
    \item validate ETag behavior with an HTTP 304 response;
    \item download one image;
    \item independently acquire and decode one benchmark image;
    \item render a synthetic 50-row list;
    \item scroll a synthetic 200-row list;
    \item write and read 200 local records or their aligned representation;
    \item upload the deterministic benchmark image.
\end{enumerate}

The ETag wrapper performs a preparatory HTTP 200 retrieval before the conditional request that returns HTTP 304. It therefore measures a two-request validation workflow. The outer full-scenario duration includes every wrapper, backend preparation, UI-driver time, and orchestration between wrappers.

\subsection{Run Protocol and Runtime Data Quality}

Each generated CSV contains 2,000 sequential wrapper observations from one in-process execution session. Native Android, native iOS, Flutter Android, Flutter iOS, React Native Android and iOS, KMP Android, and KMP iOS label them as ten batches of 200. The React Native reruns include a three-second recovery pause between runs, excluded from the measured full-scenario and operation durations.

\subsection{Repository-Metric Configuration}

The source snapshot used for RQ3 and RQ4 includes authored mobile-language files from the following roots:

\begin{itemize}
    \item native iOS: Swift under the application source tree;
    \item native Android: Kotlin under \texttt{app/src/main};
    \item Flutter: Dart under \texttt{lib} plus maintained Android Kotlin and iOS Swift host source;
    \item React Native: \texttt{App.tsx}, TypeScript under \texttt{src}, and maintained Android Kotlin and iOS Swift host source;
    \item Kotlin Multiplatform: Kotlin and SQLDelight source in \texttt{commonMain}, \texttt{androidMain}, and \texttt{iosMain}, plus the Android Compose and iOS SwiftUI shells.
\end{itemize}

Tests, build scripts, generated and build output, dependencies, lockfiles, assets, resources, caches, and exported telemetry are excluded. Physical lines of code (PLOC) count every file-line record; nonblank lines exclude whitespace-only records; the reported source lines of code (SLOC) apply a lexical filter for line and block comments, counting mixed code-and-comment lines as source. Handwritten platform host code and benchmark and telemetry code are included. The latter cannot be removed consistently because KMP intertwines benchmark UI and application-shell behavior. The counts therefore characterize the current whole client, not an instrumentation-free product.

Dependency declarations are read from each platform's maintained manifest or project configuration. Counts are reported as declared package products or runtime artifacts according to the native package manager, pub, npm, or Gradle source-set model. They describe the visible dependency surface under each ecosystem's declaration model.

\subsection{Reproducibility Materials}

The supplied result bundle contains one CSV for each of the eight runtime variants. Every file concatenates three tables: detailed nested telemetry for the last run, wrapper summaries for all 2,000 runs, and aggregate summaries for the 12 wrapper operations.

\section{Results}

The results combine the eight supplied runtime exports with a structural inventory of the five current mobile source trees. Runtime findings are descriptive because collection was sequential. 

\subsection{Dataset and Run Validity}

Table \ref{tab:run-validity} reports the observed run structure. Every variant contains 2,000 unique wrapper identifiers, every row is marked successful with 12 of 12 completed steps, and every operation aggregate reports 2,000 samples. The dataset therefore contains 16,000 completed wrapper runs.

\begin{table*}[t]
\centering
\caption{Wrapper-level validity and batching in the supplied runtime exports.}
\begin{tabular}{llrrrr}
\toprule
OS & Approach & Attempted & Completed & Included & Exported batch layout \\
\midrule
iOS & Native & 2,000 & 2,000 & 2,000 & $10\times200$ \\
iOS & Kotlin Multiplatform & 2,000 & 2,000 & 2,000 & $10\times200$ \\
iOS & React Native & 2,000 & 2,000 & 2,000 & $10\times200$ \\
iOS & Flutter & 2,000 & 2,000 & 2,000 & $10\times200$ \\
Android & Native & 2,000 & 2,000 & 2,000 & $10\times200$ \\
Android & Kotlin Multiplatform & 2,000 & 2,000 & 2,000 & $10\times200$ \\
Android & React Native & 2,000 & 2,000 & 2,000 & $10\times200$ \\
Android & Flutter & 2,000 & 2,000 & 2,000 & $10\times200$ \\
\bottomrule
\end{tabular}
\label{tab:run-validity}
\end{table*}

\subsection{RQ1: Time Behavior}

\pgfplotstableread{
Label Mean Sd
iOS-Native 1606.078 93.014
iOS-KMP 1589.136 101.181
iOS-RN 1303.077 464.556
iOS-Flutter 1872.157 343.031
Android-Native 1074.042 125.781
Android-KMP 1306.726 375.725
Android-RN 2186.435 684.013
Android-Flutter 3075.470 349.366
}\elapsedsummarydata

\pgfplotstableread{
Label Mean
iOS-Native 353.120
iOS-KMP 374.677
iOS-RN 679.319
iOS-Flutter 518.638
Android-Native 452.755
Android-KMP 504.592
Android-RN 1218.088
Android-Flutter 1547.144
}\clientsubtotaldata

\pgfplotstableread{
Step Native NativeSd KMP KMPSd RN RNSd Flutter FlutterSd
Login 95.868 21.850 91.684 17.221 238.169 100.204 175.615 158.876
Fetch 26.784 20.745 38.528 30.413 53.215 42.200 50.812 52.845
Fetch304 36.365 19.225 40.044 31.292 61.563 36.282 49.389 51.177
Download 61.118 21.484 63.221 24.982 108.294 35.957 69.778 57.801
Decode 68.960 23.402 58.804 16.986 102.886 44.129 77.359 54.035
LocalSync 15.320 0.677 7.912 2.832 20.320 1.701 30.433 6.341
Upload 48.705 13.564 74.484 52.287 94.872 69.392 65.252 60.273
}\iosstepsdata

\pgfplotstableread{
Step Native NativeSd KMP KMPSd RN RNSd Flutter FlutterSd
Login 89.234 29.217 122.276 64.405 254.406 110.554 161.609 61.439
Fetch 23.442 21.226 43.478 33.151 90.483 61.878 138.068 43.448
Fetch304 34.254 26.478 54.273 35.326 133.070 124.271 236.739 66.401
Download 94.005 45.411 95.138 51.181 321.938 225.128 283.190 79.043
Decode 98.481 36.487 99.469 51.964 224.738 130.874 297.937 81.060
LocalSync 29.188 8.847 9.287 2.006 57.469 13.995 128.760 21.835
Upload 84.151 43.923 80.671 50.722 135.984 76.332 300.841 78.933
}\androidstepsdata

\begin{figure*}[t]
\centering
\begin{tikzpicture}
\begin{axis}[
    ybar,
    width=0.98\textwidth,
    height=7.5cm,
    bar width=13pt,
    ylabel={Mean full-scenario elapsed time (ms)},
    symbolic x coords={iOS-Native,iOS-KMP,iOS-RN,iOS-Flutter,Android-Native,Android-KMP,Android-RN,Android-Flutter},
    xtick=data,
    x tick label style={rotate=35, anchor=east, font=\small},
    ymin=0,
    enlarge x limits=0.12,
    grid=major,
    error bars/y dir=both,
    error bars/y explicit
]
\addplot+[draw=black, fill=blue!45]
    table[x=Label, y=Mean, y error=Sd] {\elapsedsummarydata};
\end{axis}
\end{tikzpicture}
\caption{Full-scenario elapsed time for 2,000 wrapper runs per variant. Error bars show sample standard deviation. The total includes backend reset and seed and framework-specific UI-driver time, so it is contextual rather than the primary client-framework outcome.}
\label{fig:elapsed-summary}
\end{figure*}
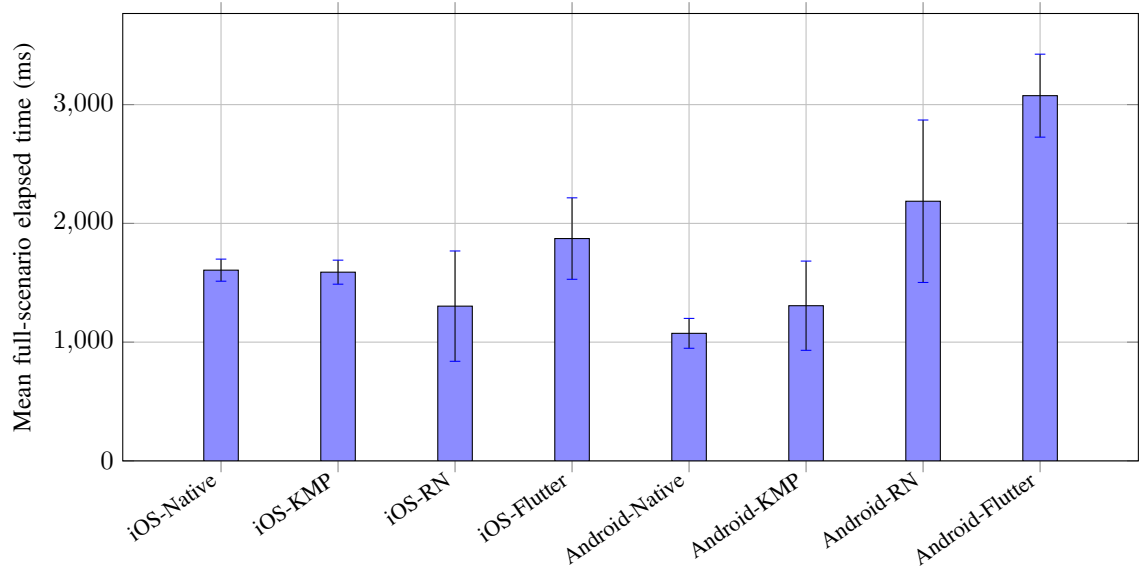

Figure \ref{fig:elapsed-summary} and Table \ref{tab:elapsed-summary} report full-scenario elapsed time. This total is contextual because it includes backend reset and seed, UI-driver time, and orchestration. Seed is not plotted as an operation in any result figure.

\begin{table*}[t]
\centering
\caption{Full-scenario elapsed-time distribution for each variant, in milliseconds.}
\begin{tabular}{llrrrrr}
\toprule
OS & Approach & Mean $\pm$ SD & Median & P95 & Range \\
\midrule
iOS & Native & $1606.078\pm93.014$ & 1585.0 & 1769.00 & 1456--2498 \\
iOS & Kotlin Multiplatform & $1589.136\pm101.181$ & 1571.0 & 1757.00 & 1415--2696 \\
iOS & React Native & $1303.077\pm464.556$ & 1201.0 & 1813.20 & 769--10971 \\
iOS & Flutter & $1872.157\pm343.031$ & 1698.0 & 2529.20 & 1428--4129 \\
Android & Native & $1074.042\pm125.781$ & 1048.0 & 1280.05 & 826--2214 \\
Android & Kotlin Multiplatform & $1306.726\pm375.725$ & 1204.0 & 1915.45 & 919--5056 \\
Android & React Native & $2186.435\pm684.013$ & 2076.5 & 3166.10 & 1224--12072 \\
Android & Flutter & $3075.470\pm349.366$ & 3022.0 & 3670.00 & 2261--9059 \\
\bottomrule
\end{tabular}
\label{tab:elapsed-summary}
\end{table*}

On Android, full-scenario ordering is Native, KMP, React Native, then Flutter. Relative to native, the means are 21.7\%, 103.6\%, and 186.3\% higher. On iOS, ordering is React Native, KMP, Native, then Flutter; relative to native, React Native is 18.9\% lower, KMP is 1.1\% lower, and Flutter is 16.6\% higher. The React Native iOS total is strongly affected by its 76.701 ms scroll wrapper, compared with 871.525 ms for native and 876.012 ms for KMP, so this ordering is not evidence of a general React Native advantage.

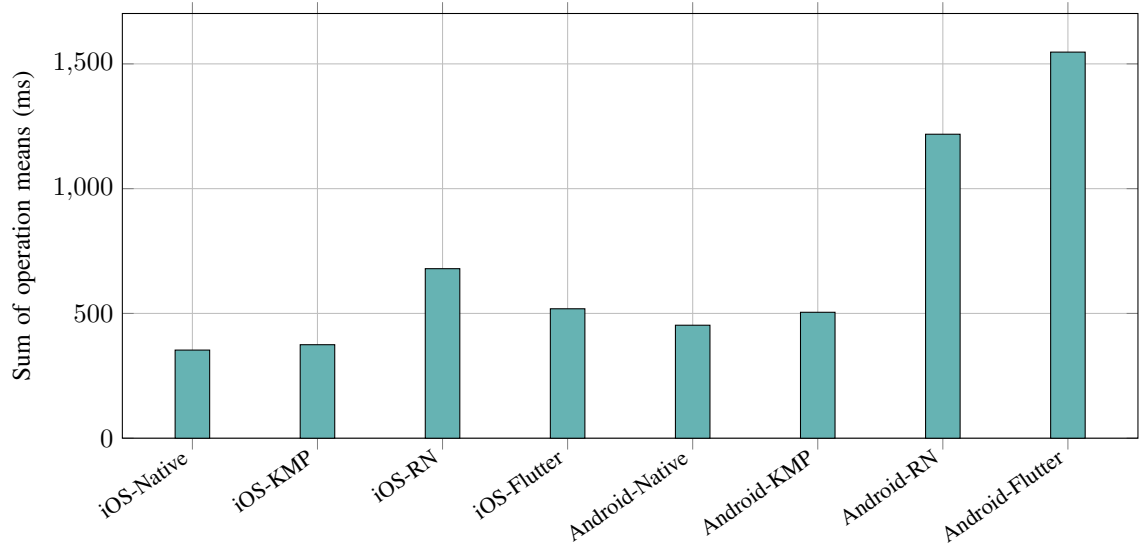
\begin{figure*}[t]
\centering
\begin{tikzpicture}
\begin{axis}[
    ybar,
    width=0.98\textwidth,
    height=7.2cm,
    bar width=13pt,
    ylabel={Sum of operation means (ms)},
    symbolic x coords={iOS-Native,iOS-KMP,iOS-RN,iOS-Flutter,Android-Native,Android-KMP,Android-RN,Android-Flutter},
    xtick=data,
    x tick label style={rotate=35, anchor=east, font=\small},
    ymin=0,
    grid=major,
    enlarge x limits=0.08
]
\addplot+[draw=black, fill=teal!60] table[x=Label, y=Mean] {\clientsubtotaldata};
\end{axis}
\end{tikzpicture}
\caption{Primary non-UI client subtotal. Bars sum aggregate means for login, fetch, conditional fetch, image download, image decode, local synchronization, and upload. Backend reset and seed and UI wrappers are excluded.}
\label{fig:client-subtotal}
\end{figure*}

Table \ref{tab:client-subtotal} and Figure \ref{fig:client-subtotal} give the primary seven-operation subtotal. It excludes the runtime marker, reset, seed, render, and scroll. Because it is calculated as a sum of aggregate means, no subtotal SD or confidence interval is reported.

\begin{table*}[t]
\centering
\caption{Seven-operation non-UI client subtotal and difference from the same-OS native subtotal.}
\begin{tabular}{llrr}
\toprule
OS & Approach & Sum of operation means (ms) & Difference from native \\
\midrule
iOS & Native & 353.120 & Reference \\
iOS & Kotlin Multiplatform & 374.677 & $+6.1\%$ \\
iOS & Flutter & 518.638 & $+46.9\%$ \\
iOS & React Native & 679.319 & $+92.4\%$ \\
Android & Native & 452.755 & Reference \\
Android & Kotlin Multiplatform & 504.592 & $+11.4\%$ \\
Android & React Native & 1218.088 & $+169.0\%$ \\
Android & Flutter & 1547.144 & $+241.7\%$ \\
\bottomrule
\end{tabular}
\label{tab:client-subtotal}
\end{table*}

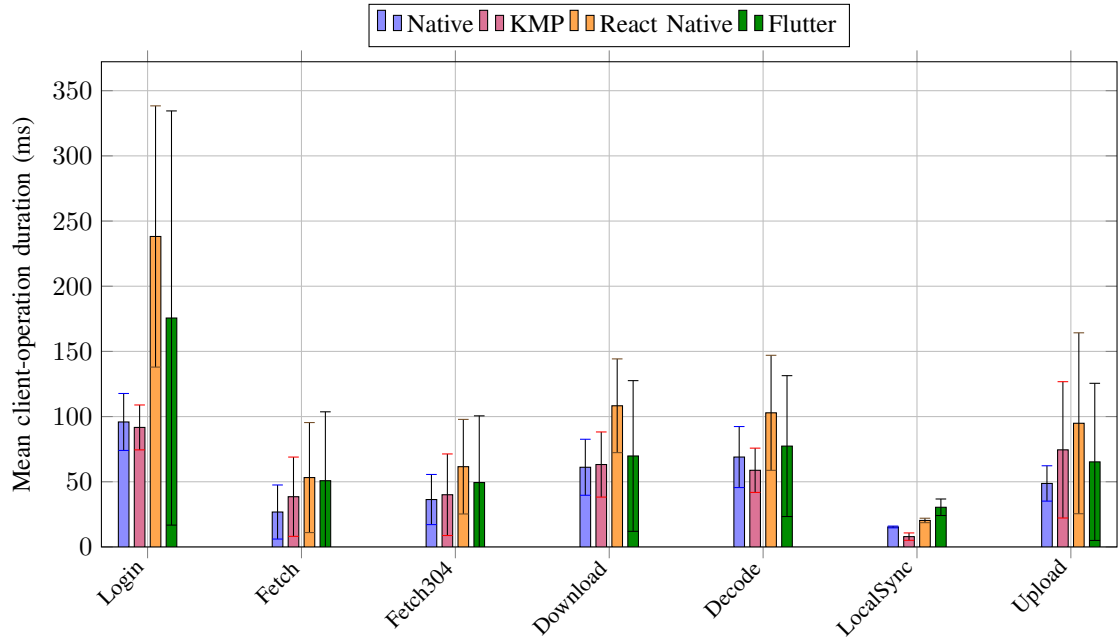
\begin{figure*}[t]
\centering
\begin{tikzpicture}
\begin{axis}[
    ybar,
    width=0.98\textwidth,
    height=8cm,
    bar width=4pt,
    ylabel={Mean client-operation duration (ms)},
    symbolic x coords={Login,Fetch,Fetch304,Download,Decode,LocalSync,Upload},
    xtick=data,
    x tick label style={rotate=45, anchor=east, font=\small},
    ymin=0,
    grid=major,
    legend style={at={(0.5,1.03)}, anchor=south, legend columns=4},
    enlarge x limits=0.05
]
\addplot+[draw=black, fill=blue!45, error bars/.cd, y dir=both, y explicit] table[x=Step, y=Native, y error=NativeSd] {\iosstepsdata};
\addplot+[draw=black, fill=purple!55, error bars/.cd, y dir=both, y explicit] table[x=Step, y=KMP, y error=KMPSd] {\iosstepsdata};
\addplot+[draw=black, fill=orange!70, error bars/.cd, y dir=both, y explicit] table[x=Step, y=RN, y error=RNSd] {\iosstepsdata};
\addplot+[draw=black, fill=green!55!black, error bars/.cd, y dir=both, y explicit] table[x=Step, y=Flutter, y error=FlutterSd] {\iosstepsdata};
\legend{Native, KMP, React Native, Flutter}
\end{axis}
\end{tikzpicture}
\caption{iOS aggregate means for the seven client operations, each based on 2,000 summary samples. Error bars show sample standard deviation. Backend reset and seed are excluded.}
\label{fig:ios-client-steps}
\end{figure*}

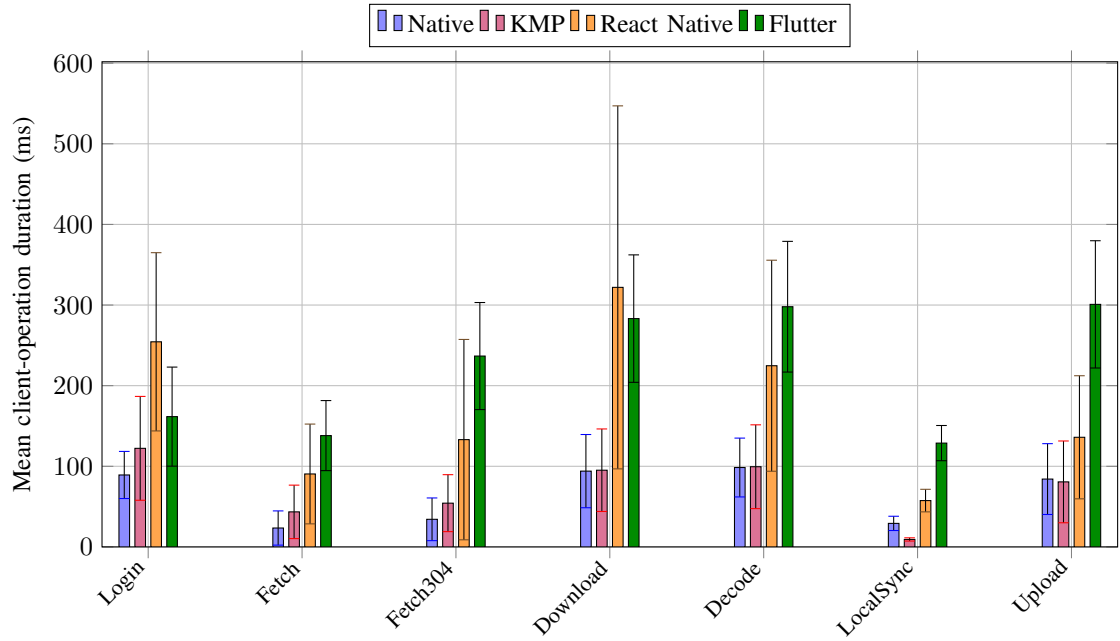
\begin{figure*}[t]
\centering
\begin{tikzpicture}
\begin{axis}[
    ybar,
    width=0.98\textwidth,
    height=8cm,
    bar width=4pt,
    ylabel={Mean client-operation duration (ms)},
    symbolic x coords={Login,Fetch,Fetch304,Download,Decode,LocalSync,Upload},
    xtick=data,
    x tick label style={rotate=45, anchor=east, font=\small},
    ymin=0,
    grid=major,
    legend style={at={(0.5,1.03)}, anchor=south, legend columns=4},
    enlarge x limits=0.05
]
\addplot+[draw=black, fill=blue!45, error bars/.cd, y dir=both, y explicit] table[x=Step, y=Native, y error=NativeSd] {\androidstepsdata};
\addplot+[draw=black, fill=purple!55, error bars/.cd, y dir=both, y explicit] table[x=Step, y=KMP, y error=KMPSd] {\androidstepsdata};
\addplot+[draw=black, fill=orange!70, error bars/.cd, y dir=both, y explicit] table[x=Step, y=RN, y error=RNSd] {\androidstepsdata};
\addplot+[draw=black, fill=green!55!black, error bars/.cd, y dir=both, y explicit] table[x=Step, y=Flutter, y error=FlutterSd] {\androidstepsdata};
\legend{Native, KMP, React Native, Flutter}
\end{axis}
\end{tikzpicture}
\caption{Android aggregate means for the seven client operations, each based on 2,000 summary samples. Error bars show sample standard deviation. Backend reset and seed are excluded.}
\label{fig:android-client-steps}
\end{figure*}

Figures \ref{fig:ios-client-steps} and \ref{fig:android-client-steps} show the component means. Reset and seed are intentionally absent because they are backend preparation, and the runtime marker and UI wrappers are analyzed separately.

The component profile is operation dependent. On iOS, KMP has the shortest login, decode, and local-sync means, while native has the shortest fetch, conditional-fetch, download, and upload means. On Android, native has the shortest login, fetch, conditional-fetch, download, and decode means, while KMP has the shortest local-sync and upload means. KMP Android is within 1.2\% of native for download and 1.0\% for decode. These observations describe the timed wrappers and retain the semantic caveats identified in the artifact audit.

\textbf{Answer to RQ1.} For the seven-operation client subtotal, native is lowest on both operating systems and KMP is the closest cross-platform implementation, at 6.1\% above native on iOS and 11.4\% above native on Android. React Native and Flutter have larger subtotals, but no implementation leads every component. Full-scenario ordering is different because backend and UI-driver work dominate parts of that total.

\subsection{RQ2: Observable Rendering Responsiveness}

Table \ref{tab:ui-wrappers} reports the available list-render and scroll wrapper summaries. Each value contains 2,000 samples.

\begin{table*}[t]
\centering
\caption{Synthetic UI wrapper durations, mean $\pm$ sample SD in milliseconds.}
\begin{tabular}{llrr}
\toprule
OS & Approach & Render 50 rows & Scroll 200 rows \\
\midrule
iOS & Native & $49.708\pm3.108$ & $871.525\pm6.199$ \\
iOS & Kotlin Multiplatform & $36.517\pm1.460$ & $876.012\pm4.958$ \\
iOS & React Native & $44.568\pm5.394$ & $76.701\pm6.606$ \\
iOS & Flutter & $33.989\pm6.014$ & $890.003\pm3.844$ \\
Android & Native & $45.338\pm5.646$ & $316.170\pm15.054$ \\
Android & Kotlin Multiplatform & $27.404\pm6.011$ & $316.347\pm9.723$ \\
Android & React Native & $102.279\pm32.224$ & $125.403\pm15.098$ \\
Android & Flutter & $95.379\pm18.814$ & $912.652\pm11.036$ \\
\bottomrule
\end{tabular}
\label{tab:ui-wrappers}
\end{table*}

The shortest recorded render wrapper is Flutter on iOS and KMP on Android. KMP is 26.5\% below native iOS and 39.6\% below native Android for this wrapper. Its scroll duration is close to the native shell on both platforms: 0.5\% above native on iOS and 0.1\% above native on Android. This proximity is expected because KMP uses the same native UI families and closely matched platform-specific drivers. React Native's much shorter scroll wrappers and Flutter Android's much longer value reveal different prescribed paths and completion rules rather than measured smoothness.

\textbf{Answer to RQ2.} The render wrapper distinguishes the realized implementations, with Flutter lowest on iOS and KMP lowest on Android. The scroll wrappers are not cross-framework responsiveness measures because the driver is part of the timed work. The current exports cannot answer which interface has better frame pacing, fewer missed frames, or better perceived responsiveness.

\subsection{RQ3: Implementation Footprint}

Table \ref{tab:source-footprint} reports the rule-based whole-client inventory defined in the methodology. SLOC is used for relative comparisons; PLOC and nonblank lines expose sensitivity to comments and whitespace. The combined-native row represents the two source trees required to deliver both operating systems.

\begin{table*}[t]
\centering
\caption{Authored mobile-source footprint in the inspected working-tree snapshot.}
\small
\begin{tabular}{lrrrrr}
\toprule
Approach & Files & PLOC & Nonblank & SLOC & Versus combined native \\
\midrule
Native iOS & 63 & 5,754 & 4,964 & 4,522 & N/A \\
Native Android & 26 & 5,030 & 4,583 & 4,583 & N/A \\
Combined native & 89 & 10,784 & 9,547 & 9,105 & Reference \\
Flutter & 16 & 6,062 & 5,467 & 5,466 & $-40.0\%$ \\
React Native & 39 & 4,681 & 4,245 & 4,209 & $-53.8\%$ \\
Kotlin Multiplatform & 16 & 4,516 & 4,196 & 4,195 & $-53.9\%$ \\
\bottomrule
\end{tabular}
\label{tab:source-footprint}
\end{table*}

KMP's 4,195 SLOC consists of 1,624 shared SLOC, 1,291 Android-specific SLOC, and 1,280 iOS-specific SLOC. Shared Kotlin therefore accounts for 38.7\% of unique KMP source. Viewed per target, common code forms 55.7\% of the Android deliverable and 55.9\% of the iOS deliverable. When common source is counted once for each target in the two-deliverable source incidence, the shared proportion is 55.8\%. These values reflect KMP's architecture of shared application services with separate Compose and SwiftUI interfaces.

\textbf{Answer to RQ3.} In this whole-client snapshot, the combined native implementation contains 9,105 SLOC. Flutter contains 5,466 SLOC, 40.0\% less than combined native; React Native contains 4,209 SLOC, 53.8\% less; and KMP contains 4,195 SLOC, 53.9\% less. KMP has the smallest measured total, 14 SLOC below React Native. The corresponding file, PLOC, and nonblank-line counts are reported in Table \ref{tab:source-footprint} under the same inventory rules.

\subsection{RQ4: Source-File Organization and Dependency Use}

Table \ref{tab:source-structure} compares all five codebases using the median and maximum SLOC among included authored files and the direct production dependencies declared by each project. The file statistics use the same inventory and lexical SLOC rules as RQ3.

\begin{table*}[t]
\centering
\caption{File-level source concentration and declared direct dependency surface.}
\begin{tabular}{lrrr}
\toprule
Approach & Median SLOC per file & Maximum file SLOC & Direct dependencies \\
\midrule
Native iOS & 43.0 & 363 & 2 \\
Native Android & 98.0 & 1,127 & 19 \\
Flutter & 130.0 & 1,989 & 10 \\
React Native & 54.0 & 498 & 10 \\
Kotlin Multiplatform & 157.5 & 1,116 & 19 \\
\bottomrule
\end{tabular}
\label{tab:source-structure}
\end{table*}

KMP has the highest median included-file size at 157.5 SLOC, followed by Flutter at 130.0 and native Android at 98.0. Flutter has the largest individual file at 1,989 SLOC. The largest native Android and KMP files contain 1,127 and 1,116 SLOC, respectively, while the React Native and native iOS maxima are 498 and 363 SLOC. The maintained declarations contain 2 external package products for native iOS, 19 native Android runtime artifacts, 10 third-party pub packages for Flutter, 10 runtime npm packages for React Native, and 19 KMP implementation declarations across common and platform source sets.

\textbf{Answer to RQ4.} The five codebases exhibit different file-level concentration profiles and declared dependency surfaces. KMP has the highest median file SLOC, Flutter has the largest included source file, and native iOS has the lowest median, maximum, and dependency count. Native Android and KMP each declare 19 direct dependencies, Flutter and React Native each declare 10, and native iOS declares 2. These are declaration counts within each ecosystem's package model.

\section{Discussion}

\subsection{Status of the Empirical Evidence}

The 16,000 completed wrapper runs provide a high-repetition descriptive basis. The central result is the same-OS seven-operation profile. Native has the lowest client subtotal on both systems and KMP is consistently the closest shared approach. The leader still changes for individual operations, showing why one total cannot describe every part of an application.

Full-scenario time tells a different story. React Native appears fastest on iOS because its scroll driver completes about 795 ms sooner than the native driver and 799 ms sooner than the KMP driver. Android KMP is 21.7\% above native in the full scenario and has a full-run coefficient of variation of 28.8\%; React Native has the largest Android coefficient at 31.3\%. iOS KMP is 1.1\% below native and has similar variability.

\subsection{Runtime and Rendering Trade-offs}

KMP's client subtotal is 6.1\% above native on iOS and 11.4\% above native on Android. On Android, its image download and decode means are within about 1\% of native, and its local-sync and upload wrappers are lower. On iOS it has the lowest login, decode, and local-sync means. These results are compatible with an architecture that shares service code but delegates HTTP engines, storage drivers, security, decoding, and UI behavior to platform implementations. They do not show that KMP itself causes the difference because implementations, libraries, and timed boundaries vary together.

The UI wrappers demonstrate a different trade-off. KMP's native shells produce scroll timings nearly identical to the corresponding native applications under closely matched drivers. Flutter and React Native use different driver durations, so their scroll totals cannot rank smoothness. The render wrapper favors Flutter on iOS and KMP on Android. Frame pacing, jank, and subjective responsiveness remain unmeasured.

\subsection{Implementation Footprint and Source-Structure Trade-offs}

All three shared approaches contain less authored mobile source than the sum of the two native clients in the current manifest. KMP and React Native are nearly equal in total SLOC, while Flutter is larger. File-level concentration differs: KMP has the highest median file SLOC, whereas Flutter contains the largest individual file. The counts describe the inspected implementations and include their benchmark and telemetry code. Dependency ecosystems also package functionality at different granularities, so declaration totals are interpreted within each ecosystem.

KMP's sharing boundary is especially important. Shared Kotlin is 38.7\% of unique KMP SLOC, while approximately 55.8\% of two-target source incidence is shared. The rest includes two native UIs and platform adapters. Flutter and React Native share their primary UI-language source, so a direct sharing-ratio comparison would conflate architectural strategy with reuse efficiency.

\subsection{Comparison with Prior Work}

The native lead in the primary client subtotal is consistent with the overall tendencies reported by Bi{\o}rn-Hansen et al., Nawrocki et al., and Oliveira et al., while the operation-specific exceptions agree with their finding that rankings depend on workload \cite{Biorn-Hansen2020EmpiricalInvestigationCrossPlatformMobileDevelopmentFrameworks,Nawrocki2021ComparisonOfNativeAndCrossPlatformFrameworksMobileApplications,Oliveira2023AnalyzingResourceUsageOverheadMobileAppDevelopmentFrameworks}. Bedogni et al. likewise show that framework and SDK combinations can reverse the ordering for individual mapping operations \cite{Bedogni2025OnLatencyPerformanceOfMobileMappingServicesTowardsVulnerableRoadUsersSafety}. The present results extend this type of comparison with KMP on both systems but remain bounded to one application and its wrappers.

Frattaroli et al. provide the closest prior five-approach application comparison, although their outcomes are package size, network traffic, and aggregate energy rather than operation timing \cite{Frattaroli2023EcologicalImpactNativevsCrossPlatformMobilePreliminaryStudy}. The current KMP footprint also supports the reuse pattern reported by De Almeida et al., Wheeler and Olszewska, and Blanco and Lucr\'edio: sharing reduces duplicated source but leaves adapters, UIs, and integration work \cite{DeAlmeida2023CrossPlatformMobileAppDevelopmentTheISCTESpotsExperience,Wheeler2022CrossPlatformMobileApplicationDevelopmentForSmartServices,Blanco2021HolisticApproachCrossPlatformSoftwareDevelopment}.

\subsection{Implications for Framework Selection}

Technology selection cannot be reduced to one end-to-end timing bar. Teams whose priority is the measured service workflow should examine the same-OS native and KMP profiles closely, while teams prioritizing maximum UI sharing may accept different runtime and integration characteristics from Flutter or React Native. KMP offers a distinct compromise: native interfaces and platform adapters combined with shared domain and data services. Source footprint, file-level organization, sharing boundaries, declared dependencies, runtime behavior, ecosystem maturity, and team expertise remain separate decision inputs.

\section{Limitations}

\subsection{Construct Validity}

Time measurements cover only time behavior, not the full ISO/IEC 25010 performance-efficiency characteristic. CPU use, memory, capacity, energy, package size, and startup are outside the current runtime dataset. The render and scroll wrappers do not measure frame time, dropped frames, jank, or subjective UX.

The seven-operation subtotal excludes backend and UI work, but its components still contain unequal parsing, image, multipart, and instrumentation behavior. It is a sum of exported operation means rather than a per-run aggregate. Full-scenario elapsed time includes reset, seed, and prescribed UI drivers. The RQ3 and RQ4 source measures characterize authored footprint, KMP's sharing boundary, file-level source concentration, and declared dependency surfaces under the stated counting rules.

\subsection{Internal Validity}

Implementation expertise, framework idioms, library selection, cache state, asynchronous work, and telemetry persistence can influence timings. Image decoders, HTTP engines, security stores, and UI drivers also differ.

Backend seed is sequential and non-transactional, and reset can leave partial state if database deletion fails after successful object deletion.

\subsection{Conclusion Validity}

Each build has 2,000 wrapper totals, but those observations are sequential and sometimes autocorrelated. The collections are not randomized or interleaved. Treating all rows as independent would therefore overstate precision.

\subsection{External Validity}

The study uses one plant-management application and one realized implementation per approach. Results may not generalize to games, real-time collaboration, media-intensive applications, offline-first systems, or background-processing workloads.

Frameworks, operating systems, compilers, libraries, and backend dependencies evolve quickly. KMP findings are specific to shared domain and data layers with native Compose and SwiftUI shells, not to every Kotlin Multiplatform design or a shared Compose UI. Flutter and React Native results are likewise tied to their selected packages and architectures. Despite these limitations, the application workflow, common backend, 16,000 wrapper executions, operation summaries, and separate runtime and codebase units provide a useful descriptive case study.

\section{Conclusions \& Future Work}

This paper compares five ChloroByte codebases and eight Android and iOS runtime variants, including a Kotlin Multiplatform implementation with shared application services and native interfaces. The supplied exports contain 2,000 completed runs per variant, or 16,000 wrapper-level runs. Separating backend preparation and UI-driver time from the primary non-UI client subtotal changes the interpretation of the results.

Native has the lowest seven-operation client subtotal on both operating systems. KMP is the closest cross-platform result at 6.1\% above native on iOS and 11.4\% above native on Android. Flutter is 46.9\% above native on iOS and 241.7\% on Android, while React Native is 92.4\% and 169.0\% above native. Individual operations have different leaders. The render wrappers favor Flutter on iOS and KMP on Android.

The whole-client source snapshot contains 9,105 SLOC for the two native applications together, compared with 5,466 for Flutter, 4,209 for React Native, and 4,195 for KMP. Shared Kotlin accounts for 38.7\% of unique KMP source and about 55.8\% of two-target source incidence. Across the five codebases, median file SLOC ranges from 43.0 for native iOS to 157.5 for KMP, while Flutter contains the largest included file at 1,989 SLOC. The inspected manifests declare 2 direct package products for native iOS, 19 runtime artifacts for native Android, 10 third-party packages for Flutter, 10 for React Native, and 19 implementation dependencies for KMP. These values describe the source organization and dependency structure of the implemented clients.

Future work can extend the workload to more devices and application domains and measure CPU, memory, energy, subjective experience, recorded development effort, and longitudinal maintenance.

\section{Acknowledgment}
The author acknowledges the use of generative AI tools, including Connected Papers, NotebookLM, Perplexity AI, and ChatGPT, in the preparation of this work. These tools were employed to assist in the refinement, rewriting, and organization of the manuscript. All ideas, analyses, and original contributions presented remain the author's own.

\FloatBarrier

\bibliographystyle{IEEEtran}
\bibliography{references}

\newpage

\end{document}